# Nomadic Non-Public Networks for 6G: Use Cases and Key Performance Indicators

Daniel Lindenschmitt[1,†], Benedikt Veith[2,∗], Khurshid Alam[3,∗], Ainur Aurembekova[4,†], Michael Gundall[5,∗], Mohammad Asif Habibi[6,†], Bin Han[7,†], Dennis Krummacker[8,∗], Philipp Rosemann[9,†] and Hans D. Schotten[10,∗†] ∗German Research Center for Artificial Intelligence (DFKI), Germany.

†Rheinland-Pfälzische Technische Universität (RPTU), Germany.

[1]daniel.lindenschmitt@rptu.de, [2]benedikt.veith@dfki.de, [3]khurshid.alam@dfki.de, [4]ainur.daurembekova@rptu.de, [5]michael.gundall@dfki.de, [6]m.asif@rptu.de, [7]bin.han@rptu.de, [8]dennis.krummacker@dfki.de, [9]philipp.rosemann@rptu.de, [10]hans_dieter.schotten@dfki.de, schotten@rptu.de

## Abstract

The landscape of wireless communication systems is evolving rapidly, with a pivotal role envisioned for dynamic network structures and self-organizing networks in upcoming technologies like the 6G mobile communications standard. This evolution is fueled by the growing demand from diverse sectors, including industry, manufacturing, agriculture, and the public sector, each with increasingly specific requirements. The establishment of non-public networks in the current 5G standard has laid a foundation, enabling independent operation within certain frequencies and local limitations, notably for Internet of Things applications. This paper explores the progression from non-public networks to nomadic non-public networks and their significance in the context of the forthcoming 6G era.
Building on existing work in dynamic network structures, non-public networks regulations, and alternative technological solutions, this paper introduces specific use cases enhanced by nomadic networks. In addition, relevant Key Performance Indicators are discussed on the basis of the presented use cases. These serve as a starting point for the definition of requirement clusters and thus for a evaluation metric of nomadic non-public networks. This work lays the groundwork for understanding the potential of nomadic non-public networks in the dynamic landscape of 6G wireless communication systems.

*Preprint.*

*Keywords—* 6G, Nomadic Networks, Organic Networks, NPN, NNPN

## 1 Introduction

Dynamic network structures and self-organizing networks will play an important role in future wireless systems, such as the 6G mobile communications standard [1]. This includes the use of energy-efficient, low-maintenance and low-support communication technology that can easily integrate all of a user's end devices into one wireless system. This demand is being driven by industry, manufacturing and other sectors such as agriculture and the public sector, which are demanding increasingly specific requirements from their systems. With the establishment of so-called non-public networks in the current 5G mobile communications standard, an important foundation stone has already been laid for the opening up of mobile communications technologies outside the traditional mobile network operators [2]. This enables users to operate an independent 5G mobile network in certain frequencies and locally limited. Unlike standard mobile networks, these are used more for Internet of Things (IoT) applications and less for traditional services such as voice transmission. The concept is basically comparable to Citizens Broadband Radio Service (CBRS) located in the frequency range from 3550 MHz to 3700 MHz. The Spectrum Access System coordinates the frequency use of CBRS-compliant devices in order to protect incumbents and priority access licensees from other CBRS users [3], [4]. On the one hand, this means that new applications can be developed that can be operated much more efficiently by using mobile communications standards or that completely new fields of deployment can be opened up. On the other hand, the increasing spread of Non-Public Networks (NPNs) is continuously improving the availability of hardware. This creates a pull effect, e.g. NPNs continue to gain in importance.

The first generation of NPNs laid an important foundation for this new area of application, which must be further developed with 6G [5]. In addition to the availability of inexpensive hardware and its simple operation, an uncomplicated and flexible approval process with regard to regulation is also a decisive aspect. However, this aspect is currently not prepared for nomadic situations [6].

In current NPNs, stationary operation continues to be the limiting factor of the network. By introducing Nomadic Non-Public Networks (NNPNs), completely new fields of application can once again be opened up and emerging 6G mobile radio technology can be used even more flexibly and comprehensively. However, the introduction of mobile components for a mobile network also brings a number of challenges with it that must first be researched and then solved. Not all use cases can be realized by nomadic networks, but within the framework of NNPNs different combination possibilities can be realized with regard to the components of a mobile radio system. These various forms of implementation can impose different demands on a new 6G network architecture. The Recommendation ITU-R M. 2160 can be seen as an important basis for standardization, which includes nomadic networks in the area of usage scenarios, especially for ubiquitous connectivity, and also introduces new metrics in the newly defined capabilities with regard to coverage and sustainability, which are necessary for the establishment of NNPNs [7], [8].

Section 2 of this paper first introduces existing work in the field of dynamic network structures, regulations for NPN operators and alternative solutions based on different technologies before Section 3 introduces use cases that can be realized or significantly improved by nomadic networks in the first place. Subsequently, specific Key Performance Indicators (KPIs) are discussed in Section 4 to evaluate these use cases and structure them into new introduced clusters of NNPNs.

## 2 Related Work

The need for Nomadic Networks (NNs) in the context of mobile radio communications firstly appeared during the 5G development phase in the context of the METIS project [9]. The presented ideas also introduce so-called nomadic nodes [10]. These are single base stations that are able to increase either capacity or coverage of a cellular network [2]. Furthermore, the nomadic nodes are tightly coupled to moving networks concept [11]. Hence, it is assumed that moving devices, such as vehicles, serve as nomadic node for the aforementioned benefits. Hence, intense investigations in this context were carried out. While [2] investigates a flexible network deployment and its operation for nomadic nodes, [12] investigates the registration process and proposes a cell selection algorithm for nomadic nodes. This is crucial because the optimal cell could change over time, as nomadic nodes only extend the range of public networks. Furthermore, in the context of novel 5G releases the idea of moving networks has been extended to Non-terrestrial networks (NTNs), so that not only fixed base stations serve as aggregation point, but also Unmanned Aerial Vehicles (UAVs) and satellites [13]. In addition, the possibility of private and self-operated cellular networks was introduced [3]. This paves the way for only temporarily available networks that can be used for public events, e.g. sports or concerts [14], or emergency situations, e.g. natural or man-made disasters [15]. For latter, UAVs play an important role [16]. To integrate all aforementioned isolated solutions, NNs have to meet multiple requirements that vary upon time and location. Since this adoption capability is essential, pertinent research is in progress. The research in this direction is referred to as agile networks or organic networks. In this context a high level architecture for organic 6G networks has been developed [5]. Furthermore, an organic 6G network is assumed to be software-centric [5], whereas the from the IT domain known term programmability can also be found in this area. This leads to the idea of the integration of IT technologies into the 6G context [17] that already started in 5G by applying service-based architecture. In nomadic 6G networks, especially the demand of flexibility, is a must [18]. Furthermore, NNs need to react to external requirements to the network. This leads to the need for interactions with others through the spatial variableness, and the network's capability to be reconfigured during its lifetime. [1] estimates that the need for reconfigurability, because of desired flexibility in various dimensions, is a development in general for upcoming interconnection technologies. As NNs are spatially mobile per definition, they raise a new scenario, not required to consider with traditional mobile communications systems, which is co-existence. In this context, so-called underlayer networks are a well-suited concept for NNs. Here, the network-in-networks is applied. Therefore, devices of moving or temporary NNs can use resources of different networks, such as existing factory networks [19]. In order to realize this concept, trust between the involved devices and networks, which can be operated by multiple parties, is necessary [20]. Here, [21] tackles that use case as it introduces two core services that enables negotiation between distinct infrastructures for that purpose, while utilizing Distributed Ledger Technology (DLT) for mitigating trust issues.

## 3 Scenarios of nomadic mobile communication networks

Radio communication-based systems for data exchange play a pivotal role in a vast array of domains. The utilization of a wireless interface to convey a spectrum of information, ranging from measured values obtained from sensors to control commands for machines, consistently furnishes the essential flexibility within the overarching system. This inherent adaptability positions such systems as pivotal enablers for the seamless integration of the IoT and its diverse applications. In particular, wireless data transmission emerges as a dynamic solution adept at effortlessly accommodating the evolving demands of interconnected and autonomous systems across various sectors, including but not limited to manufacturing, agriculture, and the media industry.

## 3.1 Aspects of Cellular Non-Public Networks

The digitalization serves as a catalyst, continually ushering in novel possibilities and fostering adaptable concepts. This transformative process, fueled by the omnipresent influence of the IoT, is further propelled by the escalating accessibility of extensive data within increasingly compressed time frames. This confluence of factors not only underscores the indispensable role of wireless communication but also positions it at the forefront of shaping a future characterized by responsive connectivity and innovative solutions. The synergy between advancing digitalization and wireless data transmission establishes a foundation upon which diverse industries can build, propelling them into a realm of heightened efficiency, interconnections, and adaptability to meet the challenges of our rapidly evolving technological landscape.

The introduction of the so-called NPN has ushered in a transformative phase, expanding the horizons of wireless data exchange within this domain through a cellular communication system built on the 5G standard. This innovative development not only represents a leap forward in technology but also offers private users opportunities to establish and fine-tune their own 5G-based wireless networks within specified frequencies.

This newfound flexibility allows private users to optimize their networks according to their unique requirements, marking a departure from the limitations imposed by traditional Mobile Network Operator (MNO) and their licensing models. The shift towards private 5G networks opens the door to new use cases, as the once-binding constraints related to MNO are now obsolete. Within the designated frequencies, private users have the autonomy to seek licenses for operation, granting them the privilege to utilize these frequencies on a location-specific basis. This decentralized approach not only enhances the adaptability of 5G networks but also fosters a more tailored and efficient utilization of wireless resources. As private users navigate this era of increased control and customization, it sets the stage for the emergence of innovative applications and services. NPN signifies not just a technological advancement but a paradigm shift towards a more dynamic and user-centric landscape in wireless communication. Since the introduction of this groundbreaking option, the landscape has witnessed a steady proliferation in the number of private operators of 5G networks, alongside a notable increase in hardware suppliers and system integrators [22]. The transition from the early adopter phase of this transformative technology, businesses are increasingly embracing NPN as the fundamental technologies for their communication needs. This surge in adoption brings forth a host of new requirements and use cases that surpass the capabilities of the initial NPN versions or fall outside the purview of existing standardization.

This phenomenon is particularly evident in the realm of NNPNs, which represent a distinct form of NPN. Nomadic networks distinguish themselves by overcoming the previously mandatory location-based licensing, allowing them the freedom to be relocated during operation.

Special requirements for the system architecture of new 6G networks and the protection of trustworthy data are further essential components that need to be examined in more detail. As the NPN ecosystem continues to mature, the exploration and refinement of these essential elements become an important task in ensuring the seamless integration and sustainable growth of next-generation communication networks. This newfound flexibility enables new use cases, each presenting specific advantages and challenges, both from a technical and regulatory standpoint. For this reason, the following subsection explains promising applications in detail.

## 3.2 Use cases in the scope of NNPNs

With the incorporation of spatial mobility as an integral component and possibility within the NPN framework, a multitude of novel use cases comes to the forefront, presenting distinct advantages for both radio communication and the applications facilitated directly through spatial mobility. This innovative inclusion not only expands the scope of NPN applications but also introduces a paradigm shift in how wireless communication systems can be leveraged in dynamic and ever-changing environments. The infusion of spatial mobility opens up possibilities for enhanced connectivity in scenarios where traditional static networks might face limitations. This evolution to a NNPN concept introduces a new dimension of adaptability, enabling seamless communication in contexts where movement is inherent, such as public transportation, agriculture, or even in rapidly changing urban environments. In the following, specific use cases and their realization by means of NNPNs are introduced and explained:

### 3.2.1 PMSE

Various categories of events and news production in the area of Programme Making and Special Events (PMSE) heavily depend on radio technology. Topics like elections or events as music festivals or sports, and analogous activities aren't confined to fixed locations. Instead, they unfold at diverse venues and times as illustrated in Figure 1a). Within media production, there are pre-planned significant events like music festivals and sports gatherings, as well as unforeseen requirements, particularly in the realm of news production. News production necessitates agile and swift responses to ongoing events while concurrently satisfying the public's appetite for high-quality reporting. Essential apparatus such as cameras or microphones, and analogous devices are fundamentally operated without physical connections. Employing wireless production yields benefits in terms of accelerated setup and dismantling, increased adaptability, and expanded freedom of movement. Present media production technology exhibits proprietary solutions tailored to specific applications and the available spectrum. Given the predominantly nomadic nature of media productions, the tools used for production must be universally accessible and applicable in this context.

The demand for radio links prompts the exploration of novel technologies. 5G NPNs, designed as unified radio network with predictable Quality-of-Service (QoS) and low latency communication operated by private stakeholders, stand out as a solution for specific planned event-based scenarios involving the operation of wireless equipment. Modifying existing conditions to facilitate nomadic or adhoc networks should permit the operation of such networks for unplanned or unscheduled events, e.g., in the scope of breaking news transmission.

In the area of PMSE, this results in specific requirements where NNPN concepts can provide a suitable framework: such as simple light-weight licensing procedures, the need for short-term local, non-local licenses and possible crossborder operation. In addition, automatic coordination with other local users is essential in ideal cases [23].

### 3.2.2 PPDR

Facilities such as fire departments, ambulances, emergency teams, etc. are covered by the scope of Public prolonged and widespread disaster operations spanning days and weeks, the coordination of activities among various organizations and health services can be effectively managed through NN. Moreover, the establishment of an operational control room becomes essential for the systematic presentation of operational and sensor data originating from diverse sources. Crucial to this is the integration of localization and control functions within the operational control room. The transmission and reception of vital control and sensor data occur with minimal latencies, incorporating 360° image data in live streams and high bandwidths. NTN platforms offer a range of benefits for data collection. Equipped with specialized sensors onboard, airborne platforms can detect seismic activity, monitor gas emissions and assess environmental damage with precision and agility. Satellites can provide a broader perspective by monitoring large-scale environmental phenomena from space. Airborne platforms can offer significant advantages for monitoring, remote sensing, communication and broadcasting capabilities, as well as detailed and up-to-date information about potential hazards without the need to endanger human lives. Airborne platforms such as drones can be quickly deployed to hazardous locations providing immediate aerial surveillance and assessment. In the event of earthquakes, tsunamis, or chemical spills,

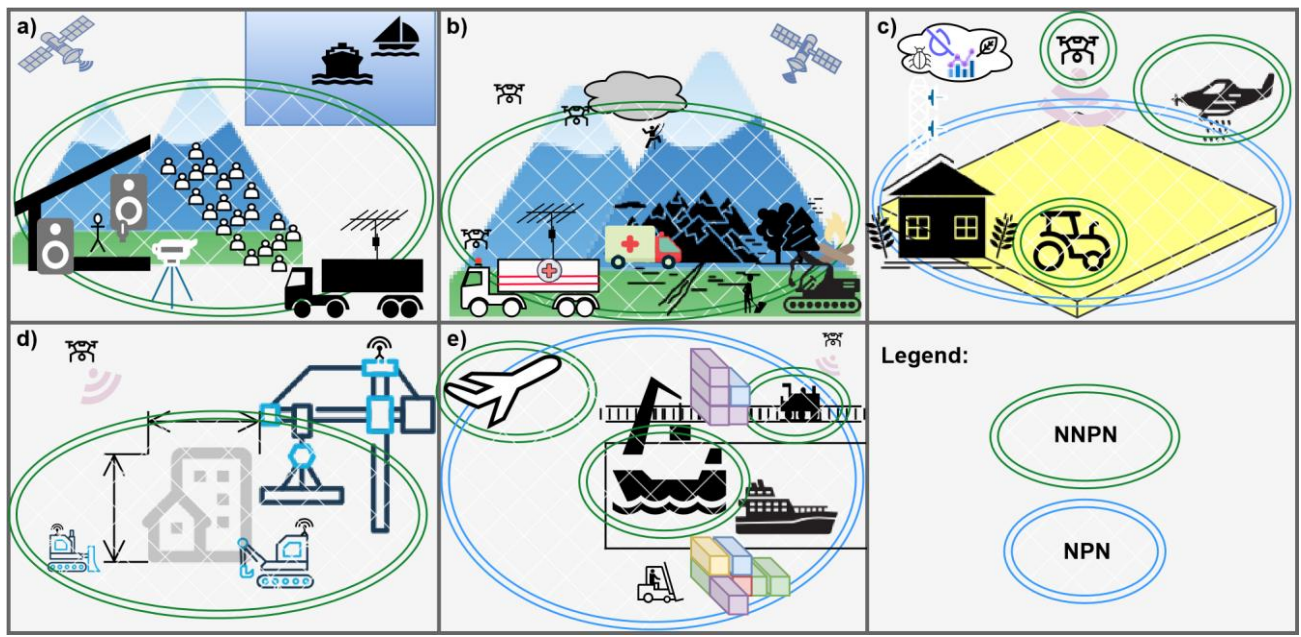

Figure 1 Possible scenarios of nomadic non-public networks: a) PMSE b) PPDR c) Agriculture d) Construction sites e) Public and private transport

Protection and Disaster Relief (PPDR). Such face distinctive and occasionally complex challenges with each operation. In operational scenarios, 5G facilitates the realization of data transmission from drones, ground-based robots, and mobile clients. As shown in Figure 1b), a large number of disaster relief scenarios such as wildfires or earthquakes or public protection situations such as demonstrations or hazardous situations can be represented more flexibly and comprehensively with NNPNs. Especially in the context of drones can be dispatched swiftly to gather critical information without delay, enabling rapid response and mitigation efforts. Equipped with high-resolution cameras, thermal imaging sensors or gas detection equipment, airborne platforms can provide real-time monitoring of dangerous manufacturing environments and create detailed 3D models of disaster-affected areas. Evaluation of data collected from ground-based, airborne and spaceborne platforms enable a comprehensive understanding of hazardous environments. In mass events such as football matches, concerts, or political protests, security agents equipped with smart glasses connected to a network can leverage remote-controlled drones equipped with cameras to enhance situational awareness. These drones provide top-down and wide-angle views, enabling real-time monitoring and analysis of crowd behavior. This technology supports effective decision-making for ensuring citizen safety, managing emergencies, and handling daily operations with increased efficiency and precision.

In order to implement the applications mentioned, instant and automatic licensing and the use of mobile wireless networks is required to fully map the scenarios mentioned above.

### 3.2.3 Agriculture

To optimize agricultural production efficiency and mitigate environmental impact, precision farming methodologies in the scope of Smart Agriculture will see increased adoption in the coming years. This entails the ongoing automation of agricultural machinery, requiring a higher degree of automation and autonomous collaboration between machines, which, in turn, necessitates local connectivity. For instance, an UAV could independently traverse a field, relaying images or videos for subsequent analysis, which is indicated in Figure 1c). During this assessment, issues such as pest infestation, heightened weed growth, water scarcity, or nutrient deficiencies might be identified. Not limited to that, UAVs have various applications in agriculture, including crop monitoring, chemical distribution, livestock health monitoring, and remote sampling. They are particularly useful for surveying large areas of land and gathering real-time data, which is essential for precision agriculture. Another area of application is the timber industry, where it also makes no economic sense to install static infrastructure everywhere in the forest. To prolong the UAV flight duration, the analysis cannot be conducted locally on the drone; it must be outsourced to the edge cloud. For prompt resolution of identified problems, data must be directly transmitted from the drone to the evaluation site. However, especially in rural areas, high-capacity public mobile networks are not yet universally available. Even where present, these networks, due to their configuration, focus in download capacity but lag in the required upload speed. Given the vast expanse of fields, Wi-Fi is unsuitable for field communication. A plausible solution for connectivity challenges could be nomadic networks, exemplified by those leveraging 5G technology like NPN. As fields are not consistently cultivated, and constructing a permanent infrastructure proves cost-prohibitive, NNs emerge as a viable solution to bridge the connectivity gap due to time restricted licensing and usage of radio spectrum also reducing the cost of adapting new technologies and ensure efficient radio resource utilization.

### 3.2.4 Construction sites

The construction industry plays an important role all around the world where a high amount of investments is concentrated. Additionally, there is a substantial demand for affordable housing that needs attention in the following years. Addressing this demand requires, among other things, the imperative of digitization and enhanced automation within the construction sector as illustrated in Figure 1d). Robotics emerges as a pivotal technology in this transformation, with its integration anticipated to bring about significant efficiency improvements to construction logistics, waste reduction, construction time minimization, safety, and quality assurance.

Cadastre and land management processes are undergoing a transformation with the integration of UAVs technology. By employing UAVs for land surveys, stakeholders gain unprecedented access to detailed insights into large swathes of terrain and their surrounding environments that allow for comprehensive assessments of topography, vegetation and infrastructure to make more accurate estimations and calculations regarding potential construction objects. The efficiency of drone surveys enables faster data acquisition, reducing the time required for planning and feasibility studies. Moreover, the data collected can be utilized beyond construction purposes, serving various land management needs such as environmental monitoring, resource management, and urban planning.

To introduce robotics effectively on construction sites and land management processes, private, nomadic, and customizable communication infrastructures are essential. Only through such infrastructures can specific requirements related to mobility, latency, jitter, confidentiality, security, and robustness be reliably fulfilled. In the scope of constructions, the changing environment needs to be considered at any time, having impact e.g. to the coverage of a network. It is in the nature of construction that new physical infrastructure is created, and the network needs to adapt on these changes. This means that the network needs to e.g. roam along a predefined path vertical and horizontally, with longterm and short-term usages. Especially in urban areas, it can be assumed that the construction site is already surrounded by public networks. The application places special demands on robots, reliability, etc., which is why a dedicated nomadic network infrastructure is necessary. This must be able to communicate with neighboring or overlapping mobile network operators due to frequency allocation, among other things.

### 3.2.5 Public and private transport

In the dynamic landscape of transportation, encompassing both passenger and freight transit, the pivotal role of networking all components within these systems is steadily gaining importance. Traditionally, specific solutions tailored for diverse application scenarios, such as those in air or rail transport and logistics, have posed challenges by restricting the fluid exchange of information and data beyond individual systems. An additional layer of complexity in this domain arises from the necessity for solutions to comply with diverse licensing and regulatory frameworks across countries and continents, a particularly intricate matter in air traffic or shipping, which is also indicated in Figure 1e). For NNs to be able to operate beyond national borders, global standardization is necessary, e.g. in the context of 6G.

Notably, NPN based on the 5G standard have already undergone successful testing and implementation in segments of the transportation sector. For instance, aircraft now receive real-time traffic and weather updates on their on-board computers while taxiing onto the runway, a feat previously achievable only through a cable connection at the gate. Nevertheless, the existing static scenarios in the NPN domain represent merely a fraction of the potential use cases, and their scope can be substantially broadened through nomadic operation. This operational flexibility facilitates the introduction of more adaptable solutions, ensuring efficient and resilient communication networks across the expansive spectrum of transportation scenarios. The use of UAVs for railway safety and the efficiency of long-distance railway inspections is a promising area of research. Studies highlight the potential of remote-operated interconnected UAVs for real-time high-resolution video transfer, which could greatly enhance the analysis and detection of traffic incidents, railway infrastructure monitoring, and inspection of railway engineering facilities, such as bridges and viaducts.

In addition, UAVs represent a transformative solution for optimizing seaport logistics based on NTN and NNPN concepts. In terms of information sharing, UAVs offer unparalleled capabilities for aerial surveillance and data collection. They can provide real-time updates on vessel arrivals and departures, monitor cargo movements and conduct security patrols along port perimeters. By conducting aerial surveys and monitoring environmental parameters such as air and water quality, UAVs help port authorities assess their ecological impact and implement mitigation measures. UAVs can serve as key enablers for digitizing port infrastructure and processes.

Another maritime application for NNPNs is on board ships. Here, NNPNs enable low-latency, self-organizing broadband connectivity between ships. All ships within a fleet form a group of independent NPNs, which are connected via an inherently variable topology. This promotes the exchange of information among each other, e.g., intrafleet communication, and enables inter-fleet communication [24].

## 4 Evaluation metrics for NNPN

### 4.1 Discussion on relevant KPIs

Considering the comprehensive list and the delineation of diverse applications and areas of use described in 3.2, a necessity emerges for a categorization of NNPNs based on core characteristics like coverage size, planability, or the number of end devices. Therefore, essential KPIs in the scope of NNPNs have to be defined. This examination leads then to the formulation of cohesive clusters within nomadic networks, effectively encapsulating the diverse set of requirements and potentials inherent in these new types of mobile networks. This is ensuring a strategic alignment with the identified needs and possibilities within the nomadic network paradigm. The varied use cases have been systematically collated in Table 1, accompanied by the introduction of pertinent KPIs. These KPIs are described in further detail below. They serve as crucial benchmarks for assessing the efficacy and adaptability of NNPN across distinct scenarios and functionalities:

Duration indicates the necessary uptime or availability of the NN. The duration can span several time units e.g. from hours to days to fulfill the demands of the application. Schedulability assesses whether the scenario can be scheduled beforehand or needs to be deployed spontaneously e.g. to unforeseen events.

Table 1 Overview of specific KPIs for NNPN demands

|  | PMSE | PPDR | Agriculture | Construction sites | Public/private transport |
|---|---|---|---|---|---|
| Duration | hours | hours to days | hours | days to months | minutes to hours |
| Schedulability | possible | no | yes | yes | probably |
| Motion predictability | yes | no | yes | possible | yes |
| Maximum coverage | micro/pico-cell | micro-cell | micro-cell | pico-cell | pico/femto-cell |
| Cross-border interaction | no | possible | no | no | possible |
| Number of active operator | multiple | multiple | single | single | multiple |
| Uplink to public network | necessary | necessary | optional | optional | optional |
| Safety-relevant communication | possible | yes | no | no | no |
| QoS aspect | eMBB | eMBB & URLLC | URLLC & mMTC | URLLC & mMTC | mMTC |
| Key aspect according to [8] | immersive communication | hyper reliable & low-latency | ubiquitous connectivity | ubiquitous connectivity | massive communication |

Motion predictability refers to the possible movement of a NN. While these networks can also be used statically, a subdivision is necessary with respect to the motion predictability of the networks in relation to their geocoordinates. This includes distinguishing whether a NN moves along a previously planned path or randomly in spatial terms.

Maximum coverage gives a statement about the required network coverage subdivided into micro-cells (less than two kilometers), pico-cells (less than 200 meters) and femtocells (less than 20 meters).

Cross-border interaction indicates whether an application may only be operated in a specific country or must be able to operate across borders.

Number of active operator provides information on whether different operators are active within the coverage area of a NN and therefore have to share existing radio resources.

Uplink to public network specifies whether the NN requires an uplink to transfer data e.g. to a central cloud server or control center.

Safety-relevant communication provides an indication of whether the use case collects and transmits safetyrelevant information, e.g. with regard to the protection of the population.

QoS aspect displays the most important aspect in relation to e.g. data rate or latency.

Key aspect according to [8] represents the most essential aspect in relation to ITU-R M.2160-0 recommendations.

## 4.2  Definition of NNPN application cluster

The detailed description of use cases and KPIs enables a further structuring of the subject area of NN from the perspective of the operators and also from the one of the network architecture. The requirements and the implications to a potential architecture should be further investigated in a next step. The effects of different KPIs on the purpose and implementation for operators and the resulting limitations can be used to introduce realization groups of NNPNs. As an outcome of Section 3 and Table 1 clusters can be defined. Each cluster covers several KPIs at the same time and has no intersection with other clusters, as introduced below:

- Unpredictable, Unscheduled, and Safety-relevant Nomadic Non-Public Network (UUS-NNPN): This cluster covers all NNs that have to react quickly to unplanned events in particular, where several operators have to act simultaneously within a coverage area. Includes PPDR and PMSE in particular.
- Scheduled, Single Operator Nomadic Non-Public Network (SS-NNPN): This category includes all NNs that can exhibit nomadic but motion predictable behavior and are used exclusively by one operator. This includes, but is not limited to, applications in agriculture and on construction sites.
- Predictable, Cross-border Nomadic Non-Public Network (PC-NNPN): This classification includes NNs that move across national borders along predictable paths. This includes, in particular, applications in the field of transportation.

The identified clusters relate to the essential differences in the requirements, meaning that not all KPIs need to be covered. In particular, the QoS and key requirements are very different in terms of their characteristics, which on the one hand demonstrates the heterogeneous orientation of NNPNs and on the other hand means that they cannot be clearly assigned to a cluster. Rather, the defined clusters should each be able to cover the requirements according to [8] in particular. Although the presence of an uplink is not always necessary from a use case perspective, it can be advantageous for individual applications, therefore each cluster should possess this capability. Furthermore, duration and coverage were not used to differentiate the clusters, as these aspects are less relevant for the technical consideration than for the implementation by the operator. As a result of this process, use cases are clearly classified according to their requirements and prerequisites are already known for use cases not yet covered.

## 5  Conclusion

This paper systematically compiles various use cases and introduces KPIs to assess the effectiveness and adaptability of NNs across different scenarios and state out the demand for spatially moving NPNs. As a first step towards this, the paper analyzes and discusses in detail different categories of use cases and their specific demands on the concept of NN. This detailed description of use cases facilitates a structured understanding of NNs from operator and network architecture perspectives. Afterwards evaluation metrics based on KPIs have been defined. These KPIs include factors like duration, schedulability, motion predictability, maximum coverage, cross-border capability, number of active operators, uplink to public network, safety-relevant communication, QoS aspects, and key aspects according to ITU-R M.2160-0 recommendations. As an outcome, clusters are proposed by this paper, namely UUS-NNPN for rapid response to unplanned events, SS-NNPN for nomadic but motion predictable behavior used by a single operator, and PCNNPNs for NNs moving across borders along predictable paths. These clusters address diverse requirements, ensuring coverage of key aspects while allowing variations in KPIs and specific use case characteristics.

For future work, the requirements of the defined KPIs and clusters should be investigated further concerning the architecture of an NNPN and, in general, its impact on the standardization towards 6G.

## Acknowledgment

The authors acknowledge the financial support by the German *Federal Ministry for Education and Research (BMBF)* within the project »Open6GHub« {16KISK003K & 16KISK004} and »6GTakeOff« {16KISK067}.